\documentclass[journal]{IEEEtran}
\usepackage{dsfont}

\usepackage{cite}
\usepackage{multirow}
\usepackage{color}
\usepackage{setspace}
\usepackage{clrscode}
\usepackage{amsthm}
\usepackage{amsmath,bm}
\usepackage{pict2e}
\usepackage{amsmath}
\usepackage{amssymb}
\usepackage{graphicx}
\usepackage{bbold}
\usepackage[ruled,vlined]{algorithm2e}
\usepackage{caption}
\usepackage{subcaption}
\usepackage{mathabx}
\usepackage{float} 
\usepackage[ps2pdf,
bookmarks=false,
bookmarksnumbered=false, 
bookmarksopen=false, 
colorlinks=false]{}

\usepackage{fancyhdr}

\DeclareMathOperator*{\argmax}{argmax}

\newcommand{\svec}{\textbf{s}}

\newcommand{\avec}{\textbf{a}}
\newcommand{\vvec}{\textbf{v}}

\newcommand{\vtilde}{\tilde{v}}

\newcommand{\stvec}{\widetilde{\textbf{s}}}
\newcommand{\pvec}{\textbf{p}}

\newcommand{\R}{\mathbb{R}}
\newcommand{\mT}{\mathcal{T}}
\newcommand{\mS}{\mathcal{S}}
\newcommand{\mR}{\mathcal{R}}
\newcommand{\mA}{\mathcal{A}}
\newcommand{\mP}{\mathcal{P}}
\newcommand{\mE}{\mathcal{E}}
\newcommand{\mD}{\mathcal{D}}

\newcommand{\E}{\mathbb{E}}
\newcommand{\xibm}{\bm{\xi}}

\allowdisplaybreaks[2]

\begin{document}

	\title{Resilient Path Planning for UAVs in Data Collection under Adversarial Attacks}
	
\author{Xueyuan Wang and M. Cenk Gursoy
	\thanks{ Xueyuan Wang is with School of Computer Science and Artificial Intelligence, Changzhou University,  Changzhou 213164, China (e-mail: xywang@cczu.edu.cn).}
	 \thanks{M. Cenk Gursoy is with the Department of Electrical Engineering and Computer Science, Syracuse University, Syracuse, NY, 13244 USA	(e-mail:  mcgursoy@syr.edu).}
}

	\maketitle
	
	\begin{abstract}
		 In this paper, we investigate jamming-resilient UAV path planning strategies for data collection in Internet of Things (IoT) networks, in which the typical UAV can learn the optimal trajectory to elude such jamming attacks.  Specifically, the typical UAV is required to collect data from multiple distributed IoT nodes under collision avoidance, mission completion deadline, and kinematic constraints in the presence of jamming attacks.  We first design a fixed ground jammer with continuous jamming attack and periodical jamming attack strategies to jam the link between the typical UAV and IoT nodes. Defensive strategies involving a reinforcement learning (RL) based virtual jammer and the adoption of higher SINR thresholds are proposed to counteract against such attacks.  Secondly, we design an intelligent UAV jammer, which utilizes the RL algorithm to choose actions based on its observation.  Then, an intelligent UAV anti-jamming strategy is constructed to deal with such attacks,  and the optimal trajectory of the typical UAV is obtained via dueling double deep Q-network (D3QN). Simulation results show that both non-intelligent and intelligent jamming attacks have significant influence on the UAV's performance, and the proposed defense strategies can recover the performance close to that in no-jammer scenarios.

	\end{abstract}

	\maketitle
	\thispagestyle{fancy}
	\fancyhead{}
	\chead{The final version of this paper has been accepted in  IEEE Transactions on Information Forensics and Security, DOI:10.1109/TIFS.2023.3266699}
	\renewcommand{\headrulewidth}{0 pt}

	\section{Introduction}

	Unmanned aerial vehicles (UAVs) or drones are aircrafts piloted by remote control or embedded computer programs	without human onboard.
	Owing to their mobility, autonomy, and flexibility, UAVs are expected to be utilized extensively in different use cases in the next decade  \cite{UAV_survey_YZeng}.	
	For example, they are  considered as critical components in  Internet of Things (IoT) scenarios \cite{UAV_syed2021survey}, in which devices often have small transmit power and may not be able to communicate over a long range \cite{UAV_cellular_MMozaffari}. In such cases, UAVs can be used to assist IoT applications on e.g.,  data gathering \cite{IoT_goudarzi2019data},  disaster mitigation and recovery \cite{IoT_alsamhi2019survey}.
	Efficient trajectory 	control enables the UAV to achieve higher network performance with limited terrestrial infrastructure  \cite{mozaffari2017mobile, UAV_coops2019four}.
	However, the broadcast nature of wireless transmissions makes the UAV-enabled wireless communication systems  vulnerable to jamming  attacks \cite{UAV_survey_YZeng,jamming_wang2018survey, jamming_duan2018anti,jamming_duo2020anti}, leading to	 one of the major and serious threats to UAV-aided communications, especially when the jammer is mobile\cite{jam_darsena2022detection}.
	In addition, the trajectory control problem in hostile environments faces the following challenges, which make the UAV path planning hard to perform: i) multiple practical constraints should be jointly considered; ii) there exists uncertainty in the information on the jammer; and iii) malicious jamming makes the environment time-varying and non-stationary, especially when intelligent and mobile jammers are considered.
	Motivated by these observations, this paper investigates the effective jamming-resilient policies to safeguard UAV-enabled data collection networks by designing the UAV trajectory under multiple practical constraints in adversarial settings.

	\subsection{Related Works}
	
	Several studies have recently addressed ground jamming attacks in UAV-enabled networks.
	Particularly, in \cite{jamming_duan2018anti}, a  received signal strength  based jammer localization algorithm is proposed to help the UAV plan its path.
	In   \cite{jamming_duo2020anti}, by exploiting the block coordinate descent (BCD) and successive convex approximation (SCA) techniques, an iterative algorithm was proposed to solve the anti-jamming three-dimensional UAV trajectory design problem.
	In \cite{jam_wu2019robust}, the authors considered multiple ground jammers in a multi-UAV path planning problem, and proposed 	two BCD  based algorithms to obtain sub-optimal solutions with the aid of slack	variables, SCA technique and S-procedure. In addition, the same method was constructed to solve the UAV path planning in a uplink communication system with turning and climbing angle constraints in \cite{jam_gao2020robust}.
	The authors in \cite{jam_wu2021uav} considered a UAV-enable relay network under malicious ground jamming attacks, BCD and SCA techniques were utilized to optimize the UAV trajectory and the transmit powers of both the UAV and the source node.
	In  \cite{jam_wang2018trajectory}, by introducing	the slack variables and leveraging the SCA technique, the authors designed a trajectory planning method to optimize 	the UAV’s 3D position, and cases with a single jammer and also with multiple jammers were discussed.
	The authors in \cite{jam_wu2021uavswarm}  investigated UAV swarm communication in the presence of jammers, an iterative algorithm was constructed based on BCD and SCA technique to optimize the UAVs' trajectories.
	
	Above mentioned works all considered ground jammers and mainly utilized traditional optimization techniques to solve the UAV path planning problem. These optimization-based methods typically require prior knowledge of the jammer, and lack the ability to adapt to different jamming environments, e.g., in which the jammer's location is changed, or jamming is performed in a time-varying fashion. 	
	To tackle this challenge, the authors in \cite{jam_han2021satellite}  developed a reinforcement learning (RL) based automatic flight control algorithm to perform UAV trajectory design in a coordinated satellite-UAV communication system in the presence of ground jamming attacks. The jammer launched attacks according to a jamming probability. However, this jammer was not smart either.
	The authors in \cite{jam_lin2019reinforcement} designed a deep Q-network (DQN) based UAV trajectory and power control scheme against attacks from a ground jammer, which could change its path and transmit power levels. But the jammer did not have an intelligent policy to adjust its condition to perform adaptive attacks.

	Due to the expanded application of UAVs, malicious jamming attacks may also come from the sky. Consequently, UAV jamming attacks have also been considered recently in the literature.
	For instance, the authors in \cite{jam_bhattacharya2010game} formulated a zero-sum pursuit-evasion game to compute optimal trajectory strategies by a team of UAVs to evade the attack of an aerial jammer on the communication channel between UAVs. 	
	The authors in \cite{jam_xu2018one} considered a Bayesian Stackelberg game to formulate the competitive relations between UAVs and an aerial jammer, where the jammer and the UAVs aim to complete their missions by selecting their optimal power control strategies.

	RL has also been utilized to obtain  solutions against aerial jamming attacks.
	For example, in \cite{jam_li2018protecting}, the authors utilized the non-cooperative game theory to propose a Q-learning based power control algorithm to obtain an adaptive policy against a smart UAV jammer, which executes multiple attack types, such as eavesdropping, jamming, and spoofing.
	In \cite{jam_xiao2017user}, a static and smart attacker, which made subjective decisions to choose the attack types was taken into account, and DQN based UAV power allocation strategy was proposed against the attack.
	In \cite{jam_gao2019anti}, the ground users aimed to  learn the optimal anti-jamming  policy to protect its communication with a ground base station. The optimal jamming trajectory and user communication trajectory were obtained via deep recurrent Q-network and  DQN, respectively.	
	In \cite{jam_li2021uav}, the authors considered a task-based anti-jamming scenario, in which a UAV swarm cooperated to detect a fixed ground target, and a cluster of UAV jammers cooperated to interfere with the area around the target. A knowledge-based RL algorithm  was proposed for the UAV swarm to learn jamming-resilient trajectories.
	In \cite{jam_liu2021uav}, the authors considered a maritime communication scheme, which applied a UAV as relay to forward the message between ships against smart aerial jamming attacks. Q-learning and dueling neutral networks were utilized to select power control policies for the jammer and the UAV, respectively.
	The authors in \cite{jam_peng2019anti} considered both a fixed jammer and a mobile UAV jammer with a fixed trajectory. A modified Q-learning algorithm based on multi-parameter programming was proposed for the UAVs to tune antenna beam to improve the overall communication quality.

	In above mentioned related works, smart aerial jamming attacks were considered. However, none of the studies considered jamming attacks in UAV  data collection networks with multiple practical constraints, e.g., collision avoidance constraint, kinematic constraint, communication constraint, flight duration constraint. 	
	An intelligent reflecting surface (IRS)-assisted UAV data collection network under malicious jamming was taken into account in \cite{ji2022energy}.   An alternating optimization based algorithm was proposed by leveraging the Dinkelbach's algorithm, SCA, and BDC method, which requires prior knowledge of the jammer information. The jammer in this work was on the ground and non-intelligent. Also, collision avoidance and kinematic constraints were not taken into account.
	Note that different networks require very different algorithm designs.

	\subsection{Contributions}	
	In this paper, different from prior studies, we consider a general noncooperative multi-UAV setting and address decentralized UAV trajectory designs for data collection in the presence of adversarial jamming attackers while also avoiding collision with other non-adversarial UAVs and considering multiple practical constraints.  The main contributions are summarized as follows:

	\begin{itemize}
		\item A practical environment involving multiple constraints is considered. In particular, a practical setting that includes a fixed/mobile jammer and multiple non-cooperative and non-adversarial UAVs is addressed. Collision avoidance, mission completion deadline, kinematic, and transmission constraints are taken into account.
		\item A fixed ground jammer is designed with both continuous jamming attack and periodical jamming attack strategies to jam the link between the typical UAV and IoT nodes. Information on the jammer (e.g., its location) and the channel is unavailable to the typical UAV.  Based on the UAV path planning algorithm proposed in 
\cite{XWang-IoT22}, defensive strategies involving virtual jammers and higher SINR thresholds are proposed  against both  attack strategies.
		\item An RL-based intelligent UAV jammer is designed, by which the jammer follows the typical UAV and injects interference.  Subsequently, an intelligent jamming-resilient  strategy is constructed, with which the optimal trajectory of the typical UAV is devised via dueling double deep Q-network (D3QN) with designed state parameterization process. Sophisticated reward functions is designed to find the balance between the motion, mission and communication performance.
	\end{itemize}
	
	We further note that the proposed anti-jamming algorithms are completely based on observable data from the environment, which is more realistic than that in previous studies that assume  the position of the jammer is fixed and known or  part of the channel information is known. In addition, practical constraints including collision avoidance, mission completion deadline and kinematic constraints are taken into account even for the intelligent UAV jammer (in addition to the typical UAV), leading to more realistic and practical models.
	
	The remainder of the paper is organized as follows: Section II provides the details of the considered system model. Section III introduces the ground jamming attack strategies and  the RL-based anti-jamming algorithm. Section IV describes the intelligent mobile jamming attack algorithm. Section V presents the details of defense algorithm against the intelligent jamming attack.   Section VI focuses on numerical and simulation results to evaluate the performance of the proposed algorithms. Finally, concluding remarks are provided in Section VII.

	\section{System Model}

	\subsection{Network}
\begin{figure}[h]
		\centering
		\includegraphics[width=0.45\textwidth]{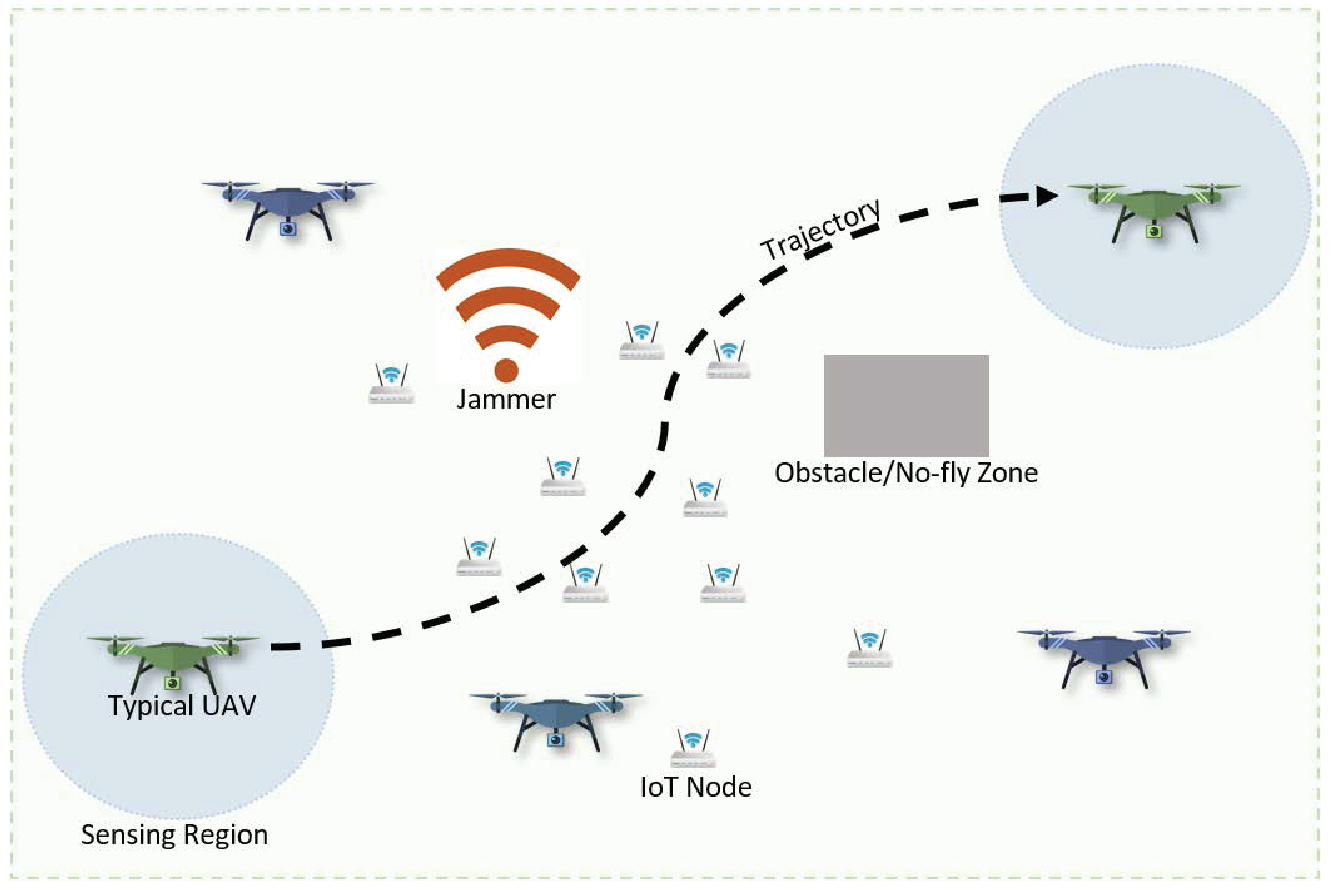}
		\caption{\small An illustration of a UAV-assisted data collection network with a jammer, which can be either a ground jammer or a moving UAV jammer. }
		\label{Fig:network}
	\end{figure}

	We assume that the area of interest is a cubic volume, which can be specified by $\mathbb{C}: \mathbb{X}\times \mathbb{Y} \times \mathbb{Z}$ and $\mathbb{X}\triangleq [x_{\min}, x_{\max}]$, $\mathbb{Y}\triangleq [y_{\min}, y_{\max}]$, and $\mathbb{Z}\triangleq [z_{\min}, z_{\max}]$.  There are multiple no-fly zones (obstacles) in the area through which UAVs cannot fly. And the no-fly zones are denoted as $\mathbb{N}:\mathbb{X}^N\times \mathbb{Y}^N \times \mathbb{Z}$. An illustration of the system model is provided in Fig. \ref{Fig:network}.
	
	\subsubsection{UAV}
	In the considered multi-UAV scenario, one UAV is chosen as the typical one, whose mission is to collect data from multiple ground IoT nodes.
	The UAV is modeled as disc-shaped with radius $r$. Let $\pvec^V = [p_x, p_y, H_V]$ denote the 3D position of the UAV, where $H_V$ is the altitude of the UAV.
	
	The typical UAV's information forms a vector that consists of the UAV's position, current velocity $\vvec^V = [v_x,v_y]$, radius $r$,  destination $\pvec^D$, maximum speed $v^V_{\max}$, and orientation $\phi$, i.e., $\svec^V= [\pvec^V, \vvec^V, r, \pvec^D, v^V_{\max},\phi]\in \R^{11}$.	
	In this multi-UAV scenario, there are also $J^o$ other non-cooperative and non-adversarial UAVs  traveling within region $\mathbb{C}$.  None of the  UAVs communicate with each other. Therefore, the missions, destinations, movements, and decision-making policies of other UAVs are unknown.
	It is assumed that the typical UAV is equipped with low-cost sensors, with which it is able to sense the existence of other UAVs when they are closer than a  certain distance.  The circular sensing region is denoted by $\mathbb{O}$.

	\subsubsection{Jammer}
	One jammer also exists in the environment, which transmits jamming signals to interfere the links between the typical UAV and the IoT nodes. The jammer can be a ground jammer with height $H_J=0$ or a moving UAV jammer with height $H_J$.
	
	\subsubsection{IoT Nodes}
	In this UAV-assisted network, there are $N$ IoT nodes that need to upload finite amount data $D^L_{n0}$ to the typical UAV via uplink transmission.   The $n^{th}$ node has transmit power $P^{n}$, and is located at ground position $\pvec^{n} = [p_{x_{n}}, p_{y_{n}}]$.
	The IoT nodes have two modes: active mode, if the node still has data to be transmitted; and silent mode, if data upload is completed.
	
	\subsection{Channel Model}
	Due to the high UAV altitude, we assume that all links between the typical UAV and IoT nodes, and link between the typical UAV and the jammer are line-of-sight (LOS). Then, the path loss can be expressed as
	\begin{align}
	L(d) =  \left(d^2 + H^2 \right) ^{\alpha/2}
	\end{align}
	where $\alpha$ is the path loss exponent, $d$ is the horizontal distance between the typical UAV and a node or a jammer, and $H$ is the height difference between the typical UAV and the IoT node (for which case we have $H=H_V$) or a jammer (for which case we have $H=|H_V-H_J|$).

	The IoT nodes and the jammer are assumed to have the omni-directional antenna gains of $G_n=0$dB and $G_J=0$dB, respectively.
	The typical UAV is assumed to be equipped with a receiver with a horizontally oriented antenna, and a simple analytical approximation for antenna gain provided by UAV can be expressed as \cite{UAV_Antenna_JChen}
	\begin{align}
	G_V (d) = \sin(\theta) = \frac{H}{\sqrt{d^2 + (H)^2}}
	\end{align}
	where $\theta$ is elevation angle between the UAV and a node. We note that even though specific antenna gains are considered for the sake of being concrete, the subsequent analysis is applicable to any type of antenna pattern.
		
	\subsection{Signal-to-Interference-plus-Noise Ratio (SINR)}
	The received signal from the $n^{th}$ node to the typical UAV can be expressed as $P^r_n = P^n G_{V}(d_{n}) L^{-1}(d_{n})$. With this, the SINR at the UAV if it is communicating with the $n^{th}$ IoT node can be formulated as
	\begin{align}
	&S^V_n \triangleq \frac{P^n  G_{V}(d_{n}) L^{-1}(d_{n}) }{\mathcal{N}_s  + I^J}
	\label{Eq:SNR}
	\end{align}
	where $\mathcal{N}_s $ is the noise power, and $I^J$ is the interference from the jammer, which can be expressed as
	\begin{align}
		I^J = P^J G_V(d_{JV}) L^{-1}(d_{JV})
	\end{align}
	where $P^J$ is the transmit power  of the jammer,  and $d_{JV}$ is the horizontal distance between the UAV and the jammer.
	
	\subsection{Rate}
	The maximum achievable information rate if the typical UAV is connected with the $n^{th}$ node is
	\begin{align}
	R^{\max}_n = \log_2(1+ S^V_n).
	\end{align}	
	To support  data flows, UAV has to maintain a reliable communication link to the IoT nodes. To achieve this, it is assumed that the SINR at the UAV when connected with a node should be larger than a certain threshold $\mT^{V}_s$. Then, the UAV can communicate with the node successfully. Otherwise, the UAV is not able to collect data from the node.
	Therefore,  the effective information rate according to the SINR threshold $\mathcal{T}^V_s$ can be given as
	\begin{align}
	R^V_n =
	\begin{cases}
	R^{\max}_n, \qquad &\text{if } S^V_n \geq \mathcal{T}^V_s, \\
	0, & \text{otherwise}.
	\end{cases}
	\end{align}

	\subsection{Scheduling}
	Standard time-division multiple access (TDMA) model is adopted. Hence, the UAV can communicate with at most one node at each time. Using $q^V_n\in\{0,1\}$ to indicate the connection with the $n^{th}$ node, we have
	\begin{align}
	\label{Eq:const_TDMA}\sum_{n}^{N} q^V_n \leq 1.
	\end{align}	
	The scheduling is according to the largest SINR strategy, meaning that  the UAV is connected with the active node providing the largest $S^V_n$. We can mathematically express the scheduling strategy as
	\begin{align}
	q^V_n =
	\begin{cases}
	1,\quad \text{if } n=\argmax\limits_{n'\in \{\text{active nodes}\}} S^V_{n'} , \\
	0,\quad \text{otherwise}.
	\end{cases}	
	\end{align}
	
	A summary of the notations are provided in Table \ref{Table:notations}.
	
	\begin{table}[htbp]
		\caption{Table of Notations}
		\label{Table:notations}
		\centering
		\begin{tabular}{l|p{2.6in}}
			\hline \hline
			\footnotesize \textbf{Notations} &  \footnotesize  \textbf{ Description}   \\ \hline \hline	 		
			\footnotesize $\mathbb{C}$, $\mathbb{N}$ &\footnotesize  Area of interest; no-fly zones/obstacles  \\  \hline
			\footnotesize $\mathbb{O}$&  \footnotesize The typical UAV's sensing region \\  \hline
			\footnotesize $N$, $J^o$&\footnotesize The number of IoT nodes; the number of other non-cooperative and non-adversarial UAVs  \\  \hline
			\footnotesize $\pvec^*$&  \footnotesize Position, where $*\in \{V,J,n\}$ and $V$, $J$, $n$ stand for the typical UAV, the jammer, and  the $n^{th}$ IoT node, respectively  \\  \hline
			\footnotesize $\vvec^*,\phi^*, r^*$&  \footnotesize  Velocity, orientation, radius, where $*\in \{V,J\}$  \\  \hline
			\footnotesize $L, G_*$&  \footnotesize Path loss, and antenna gain, where $*\in \{V,J,n\}$  \\  \hline			
			\footnotesize $H_*$ & \footnotesize  Flying altitude, where $*\in \{V,J\}$  \\  \hline
			\footnotesize $D^L_n$&  \footnotesize  The amount of data left at the $n^{th}$ IoT node\\  \hline
			\footnotesize $P^*$& \footnotesize  Transmit power, where $*\in \{J,n\}$ \\  \hline
			\footnotesize $P^r_n$&  \footnotesize  Received signal power from the $n^{th}$  IoT node \\  \hline
			\footnotesize $I^J$&  \footnotesize Interference from the jammer\\  \hline
			\footnotesize $S^V_n, R^V_n$&  \footnotesize SINR and effective information rate of the typical UAV when connected with the $n^{th}$  IoT node \\  \hline
			\footnotesize $q^V_n$&  \footnotesize  Connection indicator of the typical UAV with the $n^{th}$ IoT node \\  \hline
			\footnotesize $\mS^*,\mA^*, \mR^*$&\footnotesize State space, action space, reward, where $*\in \{V,J\}$ \\  \hline	
			\footnotesize $T^*$&\footnotesize Number of total time steps, where $*\in \{V,J\}$ \\  \hline
			\footnotesize $\mT^V_s$&\footnotesize SINR threshold of the typical UAV\\  \hline
			\footnotesize $\mT^*_t$&\footnotesize Mission completion time threshold, where $*\in \{V,J\}$ \\  \hline
			\footnotesize $\mT^*_r$&\footnotesize Maximum rotation angle in unit time duration, where $*\in \{V,J\}$ \\  \hline
			\footnotesize $v^*_{\max}$&\footnotesize Maximum speed, where $*\in \{V,J\}$ \\  \hline
			\footnotesize $\mathcal{N}_s$& \footnotesize The noise power \\  \hline			
			\footnotesize $\Delta t$&\footnotesize One time step duration  \\  \hline
			\footnotesize $\pi^*$&\footnotesize Policy, where $*\in \{V,J\}$  \\  \hline
			\footnotesize $\tau$&\footnotesize The jamming period in periodic jamming attack strategy  \\  \hline \hline		
		\end{tabular}
	\end{table}\normalsize

	\section{Ground Jamming Attacks and Defenses}
	In this section, we consider a network with a fixed ground jammer, which is  located on the ground at $\pvec^J$ and is assumed to have  transmit power $P^J$ and  omni-directional antenna pattern with $G_J=1 = 0$dB.   Therefore, the  interference from the jammer to the typical UAV can be expressed as $I^J = P^J\left(d_{JV}^2 + H_V^2 \right) ^{-\alpha/2}\frac{ H_V }{\sqrt{d_{JV}^2 + H_V^2}}$.
    Different non-learning based attack strategies are designed for the jammer to jam the links between the typical UAV and the IoT nodes,  and different defense strategies are also designed for the typical UAV  against these jamming attacks.

	\subsection{Ground Jamming Attack Strategies}

	\subsubsection{Continuous Jamming Attack Strategy}
	It is designed that the jammer transmits at a fixed transmit power $P^J_{l}$ at a fixed location all the time.
	\subsubsection{Periodic Jamming Attack Strategy}
	It is designed that the jammer works periodically at a fixed location with relatively higher transmit power $P^J_{h}$, and the jamming time duration is  $\tau^J_{h}$ seconds per minute. For fairness in the comparison with the continuous attack strategy,  it is assumed that  $P^J_{l} \times \tau^J_{l} = P^J_{h} \times \tau^J_{h}$, and $\tau^J_{l}=60$s.
	
	Note that the jammer's information and strategy are unknown to the typical UAV.
	
	\subsection{Problem Formulation for Defense} \label{subsec:typicalUAV_problemformulation}
	The goal of the typical UAV is to design efficient trajectories to maximize the collected data from the IoT nodes under several constraints in the presence of jamming attacks.  Specifically, the optimization problem can be formulated as
	\begin{align}
	(\text{PV}):\argmax_{ \{\pvec^V_t,  \forall t\}}  & \qquad \sum_{t=0}^{T^V}\sum_{n=1}^{N} q^V_{nt}\Delta t R^V_{nt} \notag  \\
	s.t. \quad
	&\label{Eq:const_collision} ||\pvec^V_t - \pvec_{jt}  ||_2 > r^V+r_j, \forall j , \forall t \tag{PV.a}\\	
	&\label{Eq:const_mission_time} T^V\cdot \Delta t \leq \mT^V_t \tag{PV.b}\\	
	&\label{Eq:const_kitc} v^V_{s_{t}} \leq v^V_{\max}, \forall t \tag{PV.c}\\
	&\label{Eq:const_kitc_angl} |\phi^V_t- \phi^V_{t-1}| \leq \Delta t \cdot \mT^V_r, \forall t \tag{PV.d}\\
	&\label{Eq:TDMA} \sum_{n}^{N} q^V_n \leq 1, \forall t \tag{PV.e}\\
	&\label{Eq:const_start_goal}\pvec^V_{0} = \pvec^S_V, \pvec^V_{T} = \pvec^{D}_V, \tag{PV.f}\\
	&\label{Eq:const_obstacle} \pvec^V_t \notin \mathbb{N}, \forall t, \tag{PV.g}
	\end{align}
	where $\pvec^V_t$ is the typical UAV's position at $t$. In the above formulation, we have collision avoidance constraints in (\ref{Eq:const_collision}) and (\ref{Eq:const_obstacle}), which restrict that the distance between two UAVs should be large than the sum of their radii all the time and the typical UAV should not collide with the obstacles/no-fly zones.  Mission completion deadline in (\ref{Eq:const_mission_time}) requires the typical UAV to finish its mission in allowed time duration. Kinematic constraints in (\ref{Eq:const_kitc}) and (\ref{Eq:const_kitc_angl}) show the maximum speed and maximum rotation angel in unit time duration limitations. (\ref{Eq:TDMA}) is TDMA constraint, and (\ref{Eq:const_start_goal}) indicates the start and destination locations constraint.
	
	\subsection{Reinforcement Learning Formulation} \label{subsec:proposed_alg}
	Typically, a sequential decision making problem can be formulated as a Markov decision process (MDP) \cite{RL_MIT}, which can be described by tuple $\langle \mS,\mA,\mP,\mR,\gamma \rangle $, representing the state space, action space, state-transition model, the reward function, and a discount factor that trades off the importance of the immediate and future rewards. Therefore, the trajectory optimization problem, as a sequential decision making problem, can be formulated as an MDP constructed as follows:
	\subsubsection{State Space $\mS^V$}
	The state is $\svec^{Vjn}_t = [\svec^V_{t}, \svec^o_t, \svec^n_t, s^V_{tt}]$ with the following components:
	\begin{itemize}
		\item  $\svec^V_{t}=[\pvec^V, \vvec, r, \pvec^D, v_{\max},\phi]$ is the typical UAV's full information vector  at time step $t$.
		\item $\svec_{t}^{o} = [ [p_{x_{jt}}, p_{y_{jt}},H_V,v_{x_{jt}},v_{y_{jt}},r_j]: j\in \{1,2,...,J^o_t \} ]$ is the joint information vector of observed other non-cooperative and non-adversarial  UAVs at the same height. $J^o_t \geq 0$ is the number observed other UAVs.
		\item $\svec^{n}_{t} = [\svec^{n}_{nt}: n \in \{1,...,N\} ] $ is the joint information vector of all IoT nodes. $\svec^{n}_{nt}=[\pvec^{n}, D^L_{nt}, P^r_{nt}]$ consists of  the location information $\pvec^{n}$, the amount of remaining data $D^L_{nt}$ (which can be obtained from $D^L_{n,t-1}$, $P^r_{nt}$, and the scheduling parameter $q^V_n$), and the received signal power $P^r_{nt}$ from each node.
		\item $s^V_{tt}$ is  the available time left for the given mission.
	\end{itemize}
	It's worth noting that information of the jammer is unknown.
	
	\subsubsection{Action Space $\mA^V$}
	The action $\avec^V$ is the index of each velocity in a velocity-set, which consists of permissible velocities sampled according to the kinematic constraints.

	\subsubsection{State-Transition Model $\mathcal{P}^V$}
	In an MDP, the state transition of an agent follows a Markov chain. Each agent takes action according to the current state, and then turns into next state after interacting with the environment. The transition probability distribution is related to the applied algorithm.

	\subsubsection{Reward $\mR^V$}
	The reward function of the typical UAV in the considered scenario can be expressed as
	\begin{align}
	\label{Eq:reward}
	\mR^V_t = \mR^V_{dt}+\mR^V_{ct} + \mR^V_{ot} + \mR^V_{tt} + \mR^V_{gt} +  \mR^V_{st}.
	\end{align}
	The first term $\mR^V_{dt}$ is related to the data collected from the nodes during next time duration $\Delta t$, and can be expressed as
	\begin{align}
	\label{Eq:reward_data}
	\mR^V_{dt} = \alpha_1 \times \left(\sum_{n=1}^N  D^L_{nt} - \sum_{n=1}^N  D^L_{n,t+1}\right).
	\end{align}
	$\mR^V_{ct}$ indicates the ``repulsive force" from other agents, and is introduced to encourage the typical UAV to stay further away from  others to avoid collision, It is given by
	\begin{align}
	\label{Eq:reward_collision}
	&\mR^V_{ct} = \notag \\
	&\begin{cases}
	-\alpha_2,                         & \text{if } d^V_{t_{\min}}\leq r^V+r_j,\\
	- \alpha_2\times(1- \frac{d^V_{t_{\min}}-r^V-r_j}{d_b^V} ),& 	\\
	&\hspace{-0.8in} \text{if } r^V+r_j<d^V_{t_{\min}} \leq d^V_b+r^V + r_j, \\	
	0, & \text{otherwise},
	\end{cases}
	\end{align}
	where $d^V_{t_{\min}}$ is the minimum distance from the typical UAV to other UAVs at the same height during next time duration $\Delta t$, and $d^V_b$ is a constant that denotes the distance buffer, inside which the typical UAV will receive a penalty that depends on $d^V_{t_{\min}}$.
	$\mR^V_{ot}$ is to penalize the collision with fixed obstacles or entering non-fly zones, and can be expressed as
	\begin{align}
	\label{Eq:reward_obstacle}
	\mR^V_{ot} =
	\begin{cases}
	-\alpha_3, & \text{if } \pvec^V_{t+1} \in \mathbb{N},\\
	0, & \text{otherwise}.
	\end{cases}
	\end{align}
	$\mR^V_{tt}$ represents the ``attractive force" from the destination to encourage the typical UAV to arrive at its destination within the allowed duration of time,  and can be formulated as
	\begin{align}
	\label{Eq:reward_time}
	\mR^V_{tt} =
	\begin{cases}
	\alpha_4 \times (s^V_{t,t+1} - T^{V\min}_{g,t+1}),  & \text{if } s^V_{t,t+1} < T^{V\min}_{g,t+1}, \\
	0, &\text{otherwise},
	\end{cases}
	\end{align}
	where $s^V_{t,t+1}$ is the available time left for the given mission,  $T^{V\min}_{g,t+1} = d^V_{g,t+1}/v^V_{\max}$ is the minimum time duration needed to reach destination, and $d^V_{g,t+1}$ is the distance to destination at time step $t+1$.
	$R^V_{gt}$ is the reward given for arriving at the destination, and
	\begin{align}
	\label{Eq:reward_goal}
	\mR^V_{gt} =
	\begin{cases}
	\alpha_5, & \text{if } \pvec^V_{t+1} = \pvec^D_V, \\
	0, & \text{otherwise}.
	\end{cases}
	\end{align}
	The last term $\mR^V_{st} =- \alpha_6$ is a step penalty for each movement, and it is used to encourage fast arrival. Note that $\alpha_{1 \sim 6}$ are positive constants, and can be varied to adjust the weight or emphasis of each reward term to adapt to different mission priorities.

	RL is a class of machine learning methods that can be utilized for solving sequential decision making problems with unknown state-transition dynamics \cite{chen2016decentralized}  \cite{everett2020collision}. RL can also be utilized to develop a jamming-resilient method that does not need to model the environment\cite{jam_peng2019anti}. Dueling double deep Q-network (D3QN) is a combination of dueling deep Q-network (DQN) and double deep Q-network (DDQN), and is a more effective and stable learning strategy. Thus, D3QN is used to learn the typical UAV path planning policy.
	
	\subsection{Anti-Ground-Jamming Strategies }
	Due to the uncertainty in  jammer's location, the typical UAV's policy should be trained with consideration on the influence from  jamming attacks. Therefore, two D3QN-based defense strategies are proposed as follows:
	
	\subsubsection{Defense with a Virtual Jammer in Training (VJ)}
	With this strategy, we assume that a virtual jammer exists in the environment in training, which transmits all the time at a fixed transmit power $P^{J'}$ at location $\pvec^{J'}$.
	The location of the virtual jammer can be chosen arbitrarily or randomly (and this location does not need to match the location of the real jammer, whose knowledge is assumed to be not available in training). For instance, the virtual jammer's location can be chosen according to the distribution of the IoT nodes (e.g., the geometric center of the IoT node groups). Then, a policy can be learned in this environment with virtual jammer present.
	\subsubsection{Defense with Higher SINR Threshold (HST)}
	As noted before, the transmission is reliable when the experienced SINR at the typical UAV is larger than a certain  threshold, i.e., $S^V \geq \mT^V_s$. With this defense strategy, we impose, in training, an SINR threshold $\mT^{V'}_s$ that is larger than what is needed, i.e., $\mT^{V'}_s > \mT^{V}_s$, and we have the typical UAV learn a policy using this higher SINR threshold $\mT^{V'}_s$. This leads to resiliency to increased interference inflicted by the jammer.
	
	The main algorithm is provided in Algorithm \ref{Algm:main_algm}.
		\begin{algorithm}
		\caption{UAV Path Planning Algorithm Against Ground Jamming Attacks}
		\label{Algm:main_algm}
		\LinesNumbered
		\KwIn{$\mT^{V}_s$, $\mT^{V}_t$, $v^{V}_{\max}$, $\mT^{V}_r$ }
		Initialize replay memory  $\mD$\\
		Initialize evaluation network $\xibm$ (including $\xibm^V$ and $\xibm^A$) \\
		Initialize target network $\xibm^-$ (including $\xibm^{V-}$ and $\xibm^{A-}$) by copping from $\xibm$	\\		
		$\mA^{V} \leftarrow \text{sampleActionSpact}(v^{V}_{\max},\mT^{V}_r)$ \\
		Choose defense strategy $DS$ \\
		\uIf{$DS$ is VJ}{$\pvec^{J'}\leftarrow$ randomGenerate in $\mathbb{C}$}
		\uElseIf{$DS$ is HST}{$\mT^{V'}_s \leftarrow \mT^V_s + c_s$ }
		\Else{No defense}
		\For{episode = 0: total episode $N_e$}{			
			$\mE \leftarrow \text{resetEnvironment}()$ \\
			\While{not done}{
				$\svec_{t}^{Vjn} \leftarrow \text{observeEnvironment}(\mE)$ \\
				$\stvec_{t}^{Vjn} \leftarrow \text{parameterizeState}(\svec_{t}^{Vjn})$ \\
				$c \leftarrow \text{randomSample(Uniform (0,1))}$\\
				\eIf{ $c\leq \epsilon$ }{
					$\avec_{t}^V \leftarrow \text{randomSample} (\mA^V)$ }
				{
					$\avec_{t}^V  \leftarrow \argmax\limits_{\avec^{V'}\in \mA} Q(\stvec_{t}^{jn},\avec^{V'};\xibm)$
				} 
				$\mR_{t}^{V}, \svec_{t+1}^{Vjn}\leftarrow \text{executeAction}(\avec_{t}^{V}, \pvec^{J'}$ or $\mT^{V'}_s )$	\\
				$\stvec_{t+1}^{Vjn} \leftarrow \text{parameterizeState}(\svec_{t+1}^{Vjn})$ \\	
				Uptate $D$ with tuple $(\stvec_{t}^{Vjn},\avec_t^{V}, \mR_t^{V},\stvec_{t+1}^{Vjn})$	\\
				Sample a minibatch of $N_b$ tuples $(\svec,\avec,\mR,\svec') \sim \text{Uniform}(D)$\\
				\For{each tuple $j$}{
					Calculate target
					\vspace{-0.1in}\begin{align}
					&\hspace{0in}y_j = \notag \\
					&\begin{cases}
					\mR, \hspace{1.3in}\text{if $\svec'$ is terminal,} \\
					\mR +\gamma Q(\svec', \argmax\limits_{\avec'} Q(\svec', \avec';\xibm); \xibm^-),\hspace{0.05in}\text{o.w.}
					\end{cases} \notag
					\end{align}
				}
				Do a gradient descent step with loss $E[(y_j - Q(\svec,\avec;\xibm))^2]$ \\
				Update $\xibm^- \leftarrow \xibm$ every $N_r$ steps
			}	
		}
		\Return $\xi$
	\end{algorithm}		

	\section{Intelligent Mobile Jamming Attack}
	In this section, we design an intelligent UAV jammer to jam the transmission between the typical UAV and the IoT nodes based on the observations from the environment.

	\subsection{UAV Jammer Model}
	In this setting, an intelligent UAV jammer is assumed to have transmit power $P^J$, height $H_J$, and certain departure and landing points.  The jammer is equipped with sensors (e.g., radar) in order to sense nearby UAVs  and track the typical UAV\footnote{Note that this assumption can be realized in practice by equipping with low-cost sensors and radars.}. The jammer is further assumed to be able to eavesdrop/learn\footnote{Note that the jammer, which is able to obtain these information, is a strong adversary, and consequently makes the defense more difficult. If the typical UAV can defend against this strong jammer, it can defend other jammers better. Therefore, this work considers the worst-case scenario and provides the corresponding defense strategies.}: 1) location information of the active ground nodes assigned to the typical UAV; and 2) the typical UAV's continuous reference signal received power (RSRP) and reference signal received quality (RSRQ) reports.
	
	If the jammer travels at the same height, it needs to avoid collision with the typical UAV while trying to get close to the UAV to increase the interference. In addition, if the sine antenna pattern of the typical UAV is adopted\footnote{Note that other antenna patterns can also be utilized, only leading to different formulation of the interference $I^J$.}, it will not receive interference from the jammer (or the interference is really small) if the jammer travels at exactly the same height as the typical UAV. With this consideration, a strong jammer is designed to fly at a different height compared to the typical UAV.
	Then, the interference from the jammer can be formulated as
	\begin{align}
	I^J &= P^J G_V(d_{JV}) (d_{JV}^{2}+(H_V- H_J)^{2})^{-\frac{\alpha}{2}} \notag \\
	&= P^J|H_V-H_J|(d_{JV}^{2}+(H_V- H_J)^{2})^{-\frac{\alpha+1}{2}}
	\end{align}
	where $H_V$ and $H_J$ are the heights of the typical UAV and the jammer, respectively. 

	\subsection{Problem Formulation  for Intelligent Attack}
	The objective of the jammer is to reduce the SINR of the typical UAV  subject to collision avoidance constraints, maximum travel time constraint, kinematic constraints and the start and destination constraints, similar to what has been described in  Section \ref{subsec:typicalUAV_problemformulation} for the typical UAV. We can formulate the optimization problem as
	\begin{align}
	(\text{PJ}):  \argmax_{ \{\pvec^J_{t},  \forall t\}}  & \qquad \E \left[\sum_{t=0}^{T^J}\sum_{n=1}^{N} q^V_{nt} \frac{1}{S^V_{nt}} \bigg| \pi^V \right] \notag  \\
	s.t.  \quad
	&\label{Eq:const_collision_jammer} ||\pvec^J_t - \pvec_{jt}  ||_2 > r^J+r_j, \forall j , \forall t \tag{PJ.a}\\	
	&\label{Eq:const_mission_time_jammer} T^J\cdot \Delta t \leq \mT^J_t \tag{PJ.b}\\	
	&\label{Eq:const_kitc_jammer} v^J_{s_{t}} \leq v^J_{\max}, \forall t \tag{PJ.c}\\
	&\label{Eq:const_kitc_angl_jammer} |\phi^J_t- \phi^J_{t-1}| \leq \Delta t \cdot \mT^J_r, \forall t \tag{PJ.d}\\
	&\label{Eq:const_start_goal_jammer}\pvec^J_{0} = \pvec^S_J, \pvec^J_{T} = \pvec^{D}_J, \tag{PJ.e}\\
	&\label{Eq:const_obstacle_jammer} \pvec^J_t \notin \mathbb{N}, \forall t, \tag{PJ.f}
	\end{align}
	where the expectation $\E[]$ in the objective function is with respect to the typical UAV's decision making policy $\pi^V$, $\pvec^J_{t}$ is the position of the jammer at $t$, and $T^J$ is the total flight time of the jammer. $S^V_{nt}$ is the typical UAV's SINR if it is connected with the $n^{th}$ IoT node at $t$, and  $q^V_{nt}$ is the association indicator of the typical UAV at time step $t$. Hence, in the above optimization problem, we have collision avoidance constraints in (\ref{Eq:const_collision_jammer}) and (\ref{Eq:const_obstacle_jammer}), mission completion deadline constraint in (\ref{Eq:const_mission_time_jammer}), kinematic constraints in (\ref{Eq:const_kitc_jammer}) and (\ref{Eq:const_kitc_angl_jammer}), and start and destination locations constraint in (\ref{Eq:const_start_goal_jammer}) for the intelligent UAV jammer.

	\subsection{Reinforcement Learning Formulation}
	The problem of trajectory design for the intelligent UAV jammer is also a sequential decision making problem, and thus can be formulated as an MDP and solved via RL. The tuple $\langle \mS, \mA,  \mR \rangle$ is formulated below.
	\subsubsection{State Space $\mS^J$}	
	In this network, the jammer can obtain the following information vectors:
	\begin{itemize}
		\item  The full information vector of itself, $\svec^J_{t} = [p^J_x,p^J_y,H_J, v^J_x,v^J_y,r^J,p^J_{gx},p^J_{gy},H_J, v^J_{\max}, \phi^J]$, at time step $t$.
		\item The observable information vector of other UAVs at $H_J$ in its sensing region, i.e., $\svec_{t}^{Jo} = [ [p_{x_{jt}}, p_{y_{jt}},H_J,v_{x_{jt}}, v_{y_{jt}},r_{j}]: j\in \{1,2,...,J^{Jo}_t \} ]$.
		\item The typical UAV's observable state  $\svec^V_{t}=[p^V_{x_{t}},p^V_{y_{t}},H_V,v^V_{x_{t}},v^V_{y_{t}}, r^V]$.
		\item The location information of active IoT nodes, i.e., $\svec^{Jn}_{t} = [[p^n_{x_n}, p^n_{y_n}]: n \in \{1,...,N^{c}\} ]$.		
		\item  The available time left for the jammer, $s^J_{tt}$.
	\end{itemize}

    The observed information vectors can be   parameterized by following process:
    \begin{itemize}
    	\item The first two information vectors are transformed into jammer-centric coordinates, in which the jammer's current location is the origin and the direction to the jammer's destination is the $x$-axis, i.e.,
    	\begin{align}
    	&\stvec^J_t = [\vtilde^J_{x_t}, \vtilde^J_{y_t}, \tilde{p}^J_{gx_t}, \tilde{p}^J_{gy_t}, d^J_{g_t}, a^J_{g_t}, r^J, v^J_{\max},\theta^J_t] \notag \\
    	&\stvec^{Jo}_{t} =[ [\tilde{p}_{x_{jt}}, \tilde{p}_{y_{jt}}, \vtilde_{x_{jt}}, \vtilde_{y_{jt}},d^o_{jt}, a^o_{jt}, r_j],\notag \\
    	&\hspace{1.5in} \text{ for } j\in \{1,2,...,J^{Jo}_t \}]. \notag
    	\end{align}
    	\item The information vector of the typical UAV and the IoT nodes from the past $\tau$ time steps can be parameterized and utilized to learn the typical UAV's policy, i.e.,
    	\begin{align}
    	&\stvec^V_{t} = [\tilde{p}^V_{x_{t}},\tilde{p}^V_{y_{t}},H_{VJ},\vtilde^V_{x_{t}},\vtilde^V_{y_{t}}]\notag \\
    	& \stvec^{Jn}_{nt}=[[\tilde{p}_{x_n}, \tilde{p}_{y_n}, d^V_{nt}, a^V_{nt}]:n \in \{1,...,N^{c}\}]  \notag
    	\end{align}
    	where $N^{c}$ is the number of active nodes, $d^V_{nt}, a^V_{nt}$ are the distance and azimuth angle of the $n^{th}$ IoT node with respect to the typical UAV's location, and the node's information vector in $\stvec^{n}_{nt}$ is listed in the smallest $d^V_{nt}$ to the largest order. Then, we have
    	\begin{align}
    	\stvec^{V}_{nt} = [[\stvec^V_{t'}, \stvec^{Jn}_{nt'} ], t' \in [t-\tau, t]]. \notag
    	\end{align}
    \end{itemize}
The parameterized state vector of the jammer can be jointly expressed as
	\begin{align}
	\stvec^{Jjn}_{t} = [\stvec^J_t,\stvec^{Jo}_{t},\stvec^{V}_{nt}, s^{J}_{tt}].
	\end{align}
	
	\subsubsection{Action Space $\mA^J$}
	Based on the jammer's kinematic constraints, permissible velocities can be sampled to build a velocity-set.  The jammer's action $a^J$ is the index of each velocity in the velocity-set.
	
	\subsubsection{Reward $\mR^J$}
	The reward function of the jammer is designed based on the objective function and the constraints, i.e.,
	\begin{align}
	\label{Eq:reward_jammer}
	\mR^J_t = \mR^J_{st}+\mR^J_{ct} + \mR^J_{ot} + \mR^J_{tt} + \mR^J_{gt} + \mR^J_{dt}.
	\end{align}
	The first term is the related to the SINR experienced at the typical UAV, and it can be expressed as
	\begin{align}
	\mR^J_{st} =
	\begin{cases}
	\alpha^J_1 \times \frac{1}{S^V_{t+1}} \quad &\text{if } S^V_{t+1}>S^V_b \\
	0 \quad &\text{otherwise}
	\end{cases}
	\end{align}
	where $S^V_{t+1}$ can be obtained from the typical UAV's RSRP and RSRQ reports, and $S^V_b$ is a positive constant which is smaller than the SINR threshold.
	$\mR^J_{ct},\mR^J_{ot},\mR^J_{tt},\mR^J_{gt}$ are the reward terms for collision avoidance, fixed obstacle avoidance, maximum travel time constraint, and arrival-to-the-destination goal, respectively, and are similar to the reward terms in (\ref{Eq:reward_collision}), (\ref{Eq:reward_obstacle}), (\ref{Eq:reward_time}) and (\ref{Eq:reward_goal}), respectively.
	The last term,	$\mR^J_{dt}$, is a reward term based on the distance between the jammer and the typical UAV, and is formulated as
	\begin{align}
	\mR^J_{dt} = d_{JV_t} - d_{JV_{t+1}}.
	\end{align}

	\subsection{Intelligent Jamming Attack Algorithm}
	The jammer's action space is sampled to be discrete, and thus Q value based RL algorithms, e.g., DQN, DDQN, D3QN, can be used to learn its policy. Since D3QN is more effective and stable, we choose D3QN to learn a strong jammer policy. The training procedure can be performed using Algorithm \ref{Algm:main_algm} by eliminating lines 5-11 and utilizing the designed $\mS^J, \mA^J,  \mR^J$ in Section IV-C.

	\section{Defense Against Intelligent Jamming Attack}
	In this section, we aim to design a defense algorithm against the intelligent jamming attacks.

	\subsection{Reinforcement Learning Formulation}
	The goal of the typical UAV is to maximize the collected data from all IoT nodes in the presence of  intelligent jamming attacks, and the objective function is the same as  in (PV). Due to the jammer's existence, state space $\mS^V$ and reward function $\mR^V$ described in Section III-C should be updated correspondingly.
	\subsubsection{State Space $\mS^V$}
	Since the jammer injects interference and is generally close to the typical UAV, and the typical UAV is able to observe nearby UAVs in its sensing region, we assume that the typical UAV is able to detect the jammer all the time\footnote{The assumption can be removed. Discussions are provided in Section VI-C-3).}. Therefore, the location information of the jammer, $\pvec^J_t$, can be obtained by the typical UAV. The jammer's locations in the past $\tau$ time steps can be used to estimate the jammer's next movement, and therefore  we have
	\begin{align}
		\stvec^{J}_{t} = [\pvec^J_{t'}, t' \in [t-\tau,t]]. \notag
	\end{align}
	The observed information vectors $\svec^V_{t}, \svec^o_t, \svec^n_t$ (described in Section III-C) can be transformed into typical UAV-centric coordinates and parameterized into
	\begin{align}
	&\stvec^V_t = [\vtilde^V_{x_t}, \vtilde^V_{y_t}, \tilde{p}^V_{gx_t}, \tilde{p}^V_{gy_t}, d^V_{g_t}, a^V_{g_t}, r^V, v^V_{\max},\theta^V_t] \notag \\
	&\stvec^o_{jt} = [\tilde{p}_{x_{jt}}, \tilde{p}_{y_{jt}}, \vtilde_{x_{jt}}, \vtilde_{y_{jt}},d^o_{jt}, a^o_{jt}, r_j],\notag \\
	&\hspace{1.5in} \text{ for } j\in \{1,2,...,J^{c}\} \notag \\
	&\stvec^{n}_{nt}=[\tilde{p}_{x_{nt}}, \tilde{p}_{y_{nt}}, d^n_{nt}, a^n_{nt},D^L_{nt}, P^r_{nt}], \quad \text{ for }n \in \{1,...,N^{c}\}.   \notag
	\end{align}
	Therefore, the state of the typical UAV is updated to
	\begin{align}
	\stvec^{Vjn}_{t} = [\stvec^V_t,[\stvec^o_{jt}, j\in\{1,...,J^{c}\}],[\stvec^{n}_{nt},n\in \{1,...,N^c\}], \stvec^{J}_{t},s^V_{tt}].
	\end{align}

	\subsubsection{Reward $\mR^V$}
	To encourage the typical UAV to fly away from the jammer, an additional reward term is added to the original reward function in (\ref{Eq:reward}), which is
	\begin{align}
	&\mR^V_{Jt} =
	&\begin{cases}
	-\alpha_7 \times (1-\frac{d_{JV_{t+1}}}{d^V_{b2}}) &\text{ if } d_{JV_{t+1}}\leq d^V_{b2} \\
	0, &\text{ otherwise}
	\end{cases}	
	\end{align}
	where $d_{JV_{t+1}}$ is the distance between the typical UAV and the jammer at the next time step $t+1$, and $d^V_{b2}$ is the distance buffer which essentially defines the safe distance between the typical UAV and the jammer.
	
	\subsection{Defense Against Intelligent Attack Algorithm}
	With the modified state and reward functions, the typical UAV's policy can be retrained using modified Algorithm \ref{Algm:main_algm}.
	
	\section{Simulation Results}
	In this section, we provide simulation results to show the performance of ground/mobile jamming attack strategies and the defense strategies.
	We choose the following performance metrics:  1) success rate (SR), which is  the portion of successful trajectories among all trajectories (and a successful trajectory means that the typical UAV arrives at its destination within mission completion deadline without collisions); 2) data collection rate (DR), which is the percentage of collected data within successful trajectories; 3) arriving on time rate (TR);
	and 4) collision rate (CR), and a collision event occurs when the typical UAV collides with any of the other UAVs in the environment.
	In the figures, we use yellow areas to show the reliable transmission region, inside which the UAVs can achieve $S^V \geq \mT^V_s$. The blue triangles are the IoT nodes, and the red triangle is the jammer. The blue and green areas are the departure and landing areas, respectively. The destination of the typical UAV is denoted as a black cross in the landing area. The gray areas are the  fixed obstacles or no-fly zones. In the figures of trajectories, black-doted lines and red doted lines display the trajectories of the typical UAV and other non-cooperative and non-adversarial UAVs, respectively. The orange-doted lines depict the trajectories of the mobile jammer. The typical UAV flies at 50m, and the transmit power of the IoT node is 10 dBm ($10^{-2}$W). Other UAVs are assumed to  use optimal reciprocal collision avoidance (ORCA) \cite{van2011reciprocal} in choosing actions and determining their trajectories.
	
	The typical UAV's policy is designed as a three-layer DNN of size (256, 256, 128), and the jammer's policy is designed as a two-layer DNN of size (256, 128). In the DNNs, ReLU function is used as the activation function, and batch-normalization is used for each layer.  Adam optimizer is used to update the parameters with learning rate 0.0003.
	Batch size is set to be 256, and the regularization parameter is 0.0001.  The exploration parameter $\epsilon$ decays linearly from 0.5 to 0.1. The replay memory capacity is 1000000.

	\subsection{Continuous Jamming Attack Scenario}
	In this subsection, the jammer is located at a fixed location and transmits at a fixed power level all the time. The transmit power of the jammer is  $P^J_{l}=10^{-3}/3$W, and the SINR threshold for the typical UAV is $\mT^V_s = 3.5$.
	\subsubsection{Attack Performance}
\begin{figure*}
		\centering
		\begin{minipage}{0.32\textwidth}
			\centering
			\includegraphics[width=1\textwidth]{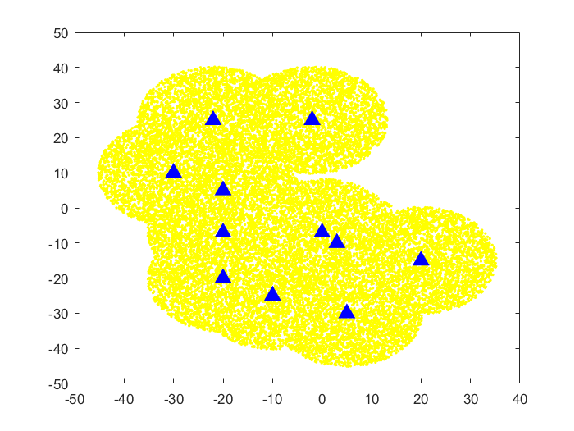}
			\subcaption{\scriptsize No jammer }
		\end{minipage}
		\begin{minipage}{0.32\textwidth}
			\centering
			\includegraphics[width=1\textwidth]{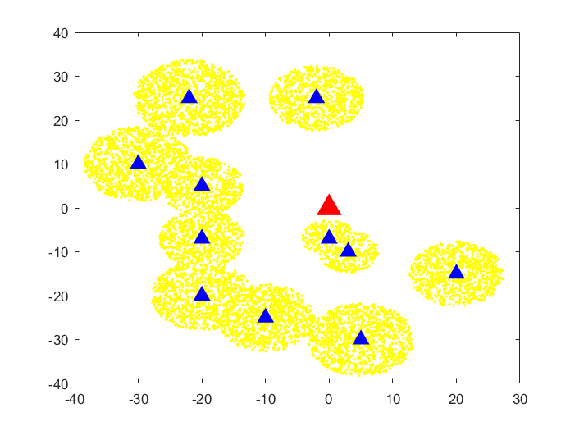}
			\subcaption{\scriptsize Jammer at (0,0)}
		\end{minipage}
		\begin{minipage}{0.32\textwidth}
			\centering
			\includegraphics[width=1\textwidth]{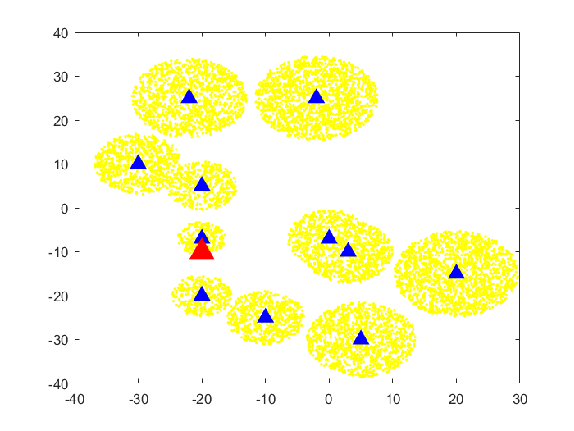}
			\subcaption{\scriptsize Jammer at (-20,-10)}
		\end{minipage}
		\caption{\small Illustrations of reliable transmission region when the jammer is absent or is located at different locations with $P^J_{l} =10^{-3}/3$W. }
		\label{Fig:fix_jammer_lower_power}
	\end{figure*}

	Fig. \ref{Fig:fix_jammer_lower_power} depicts the reliable transmission regions when the jammer is absent (in Fig. \ref{Fig:fix_jammer_lower_power}(a)) and the jammer is located at different locations (Figs. \ref{Fig:fix_jammer_lower_power}(b) and \ref{Fig:fix_jammer_lower_power}(c)). We immediately notice that the existence of the jammer significantly reduces the reliable transmission region, and different jammer  locations have varying impact.

	Table \ref{Table:fixed_jammer_attack} provides the attack performance in testing when the jammer is located at different locations $\pvec^J$ (as well as when it is absent). Note that the numerical results are averaged over 5000 testing episodes, and in each episode, the number of IoT nodes is randomly chosen from $N \in [5,10]$, the number of other UAVs is $J=2$, the locations of nodes, the start and destination points of the typical UAV, and the start and destination points of other UAVs are randomly generated.  From Table \ref{Table:fixed_jammer_attack}, we observe that the existence of the jammer substantially reduces the typical UAV's reward, SR and TR, and slightly reduces the DR. The decline in SR and TR indicates that the typical UAV needs more time to arrive at its destination. This performance degradation is due to two reasons: 1) because of the reduction in the reliable transmission region (i.e., yellow areas in the figures), the UAVs needs to get closer to each IoT node to collect data successfully, leading to a longer trajectory and longer mission completion time; and 2) the interference inflicted by the jammer changes the SINR, and this change makes the UAV get confused and not choose the optimal actions, leading to longer trajectories as well. The slight decrease in DR means that in trajectories with successful arrivals, the typical UAV can still collect  the vast majority (over 96\%) of the data in the presence of a fixed jammer.  Overall, we can state that the jammer prevents the UAV from completing its mission to a certain extent.
	\begin{table}[h]
		\centering
		\caption{Performance of continuous jamming attack.}
		\label{Table:fixed_jammer_attack}
		\begin{tabular}{c|c|c|c|c|c}
			\hline \hline
			& SR(\%) & DR(\%)   & TR(\%)   & CR(\%)  & Reward  \\ \hline
			\begin{tabular}[c]{@{}c@{}}No-\\Jammer\end{tabular} & 99.4   & 100  & 100  & 0.6 & 81.9  \\ \hline
			$\pvec^J$=(0,0)     & 84.8   & 98.1 & 85.3 & 0.5 & 18.66   \\ \hline
			(20,20)             & 95.1   & 98.7 & 96   & 0.9 & 41.89 \\ \hline
			(-20,30)           & 97.2   & 99.2 & 97.7 & 0.5 & 53.44  \\ \hline
			(-20,-10)           & 90.8   & 99.8 & 91.2 & 0.4 & 7.61   \\ \hline
			(10,-10)           & 87.2   & 96.1 & 87.8 & 0.6 & 24.83 \\ \hline \hline
		\end{tabular}
	\end{table}

	\subsubsection{Performance with Defense Utilizing a Virtual Jammer}
	Now, we deploy defensive measures and assume that, in training, a virtual jammer is located at $\pvec^{J'}=(0,0)$ on the ground with transmit power $P^{J'}=10^{-3}/3$W.  The performance results in testing (achieved by the learned policy in the presence of the virtual jammer) are presented in Table \ref{Table:fixed_jammer_defense1}.
	From this table, we observe that the SR, TR and DR can be recovered close to those in the no-jammer scenario. On the other hand, since the typical UAV needs to fly closer to the IoT nodes to get reliable connection in the presence of a jammer, the trajectories become longer. Thus, the reward with the defense strategy is still  smaller than that of the no-jammer case, since we introduce a negative reward term $R_{st}^V$ for each step (as noted at the end of Section \ref{subsec:proposed_alg}).
	
	\begin{table}[h]
		\centering
		\caption{Performance with defense strategy using a virtual jammer.}
		\label{Table:fixed_jammer_defense1}
		\begin{tabular}{c|c|c|c|c|c}
			\hline \hline
			& SR(\%) & DR(\%)   & TR(\%)   & CR(\%)  & Reward  \\ \hline
			\begin{tabular}[c]{@{}c@{}}No-\\Jammer\end{tabular} & 99.4   & 100  & 100  & 0.6 & 81.9  \\ \hline
			$\pvec^J$=(0,0)            &98.8   & 99.8 & 99 & 0.3 & 68.69   \\ \hline
			(20,20)                    &98.6 &99.9 &98.8 &0.1 &58.58 \\ \hline
			(-20,30)                   &99.1 &100 &99.6 &0.5 &73.84  \\ \hline
			(-20,-10)                  &98.7 &99.8 &99.1 &0.4 &66.8  \\ \hline
			(10,-10)                   &98.8 &99.9 &99.3 &0.5 &66.18 \\ \hline \hline
		\end{tabular}
	\end{table}

	\subsubsection{Performance with Defense Using a Higher SINR Threshold}
	With this defense strategy,  we assume that the SINR threshold is $\mT'_s = 3.9$ in training, while $\mT_s = 3.5$ in testing. The testing performance is presented in Table \ref{Table:fixed_jammer_defense2}. With this strategy, the SR, TR and DR performances can be recovered close to those of the no-jammer case, and we can observe performances similar to those of the defense with the virtual jammer.
	
	\begin{table}[h]
		\centering
		\caption{Performance with defense strategy using a higher SINR threshold.}
		\label{Table:fixed_jammer_defense2}
		\begin{tabular}{c|c|c|c|c|c}
			\hline \hline
			& SR(\%) & DR(\%)   & TR(\%)   & CR(\%)  & Reward  \\ \hline
			\begin{tabular}[c]{@{}c@{}}No-\\Jammer\end{tabular} & 99.4   & 100  & 100  & 0.6 & 81.9  \\ \hline
			$\pvec^J$=(0,0)            &97.7 &100 &99.5 &0.8 &67.46   \\ \hline
			(20,20)                    &98.2 &99.8 &98.5 &0.3 &49.49\\ \hline
			(-20,30)                   &98.8 &99.9 &99.3 &0.4 &65  \\ \hline
			(-20,-10)                  &98.8 &99.9 &99.1 &0.3 &64.83  \\ \hline
			(10,-10)                   &98.8 &99.9 &99.3 &0.4 &61.34 \\ \hline \hline
		\end{tabular}
	\end{table}

	\subsubsection{Influence of the Transmit Power Levels }
	We also present the attack and defense results in Table  \ref{Table:fixed_jammer_diff_Pt} considering different transmit powers for the IoT nodes $P^n$ and the jammer $P^J$. From Table V,  we observe that if $P^n$ is larger, the jammer expectedly needs to transmit at a high power level to have better attack performance. With the proposed strategy, i.e., defense with higher SINR  threshold, the typical UAV can successfully defend the attacks and recover the performance.
	
	\begin{table}[h]
		\centering
		\caption{Jamming attack performance and defense performance when $\pvec^J$=(0,0).}
		\label{Table:fixed_jammer_diff_Pt}
		\begin{tabular}{c|c|c|c|c}
			\hline\hline
			\multicolumn{2}{c|}{}                        & SR(\%)   & DR(\%)    & Reward \\ \hline
			\multirow{4}{*}{$P^n=10^{-1.8}$}  & No-Jammer    & 99.8 & 100  & 126.09 \\ \cline{2-5}
			& $P^J= 10^{-3}$    & 91 & 96.8  & 31.8 \\ \cline{2-5}
			& $P^J= 2*10^{-3}$    & 85.4 & 93.3  & -1.21 \\ \cline{2-5}
			& Defense &99  & 99.7  & 110.6 \\ \hline
			\multirow{4}{*}{$P^n=10^{-1.6}$} & No-Jammer   & 99.6 & 100   & 124.27 \\ \cline{2-5}
			& $P^J= 2*10^{-3}$    &90.7 & 98  & 40.9\\ \cline{2-5}
			& $P^J= 4*10^{-3}$    & 86.2  & 96.6   & 3.68 \\ \cline{2-5}
			& Defense & 98.6 &  99.8 & 106.3\\ \hline \hline
		\end{tabular}
	\end{table}
	
	\subsubsection{Trajectory Designs}
\begin{figure*}
		\centering
		\begin{minipage}{0.45\textwidth}
			\centering
			\includegraphics[width=1\textwidth]{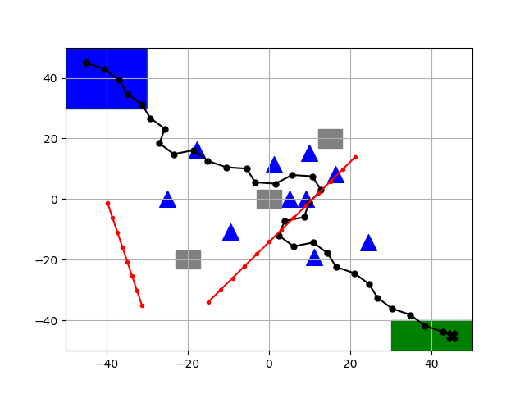}
			\subcaption{\scriptsize No jammer }
		\end{minipage}
		\begin{minipage}{0.45\textwidth}
			\centering
			\includegraphics[width=1\textwidth]{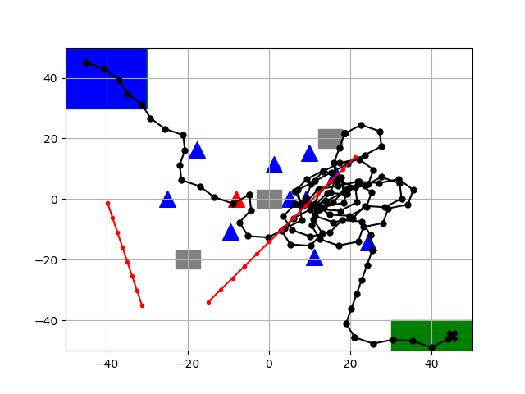}
			\subcaption{\scriptsize Continuous  attack with no defense}
		\end{minipage}
		\begin{minipage}{0.45\textwidth}
			\centering
			\includegraphics[width=1\textwidth]{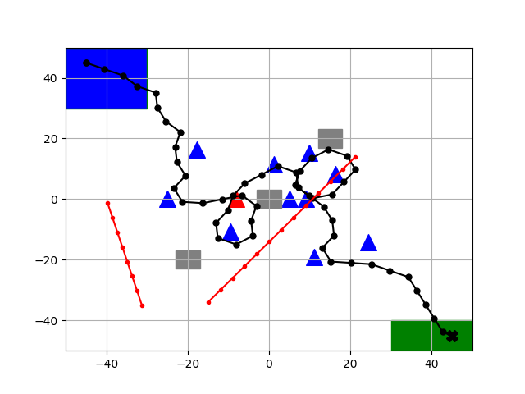}
			\subcaption{\scriptsize VJ-strategy}
		\end{minipage}
		\begin{minipage}{0.45\textwidth}
			\centering
			\includegraphics[width=1\textwidth]{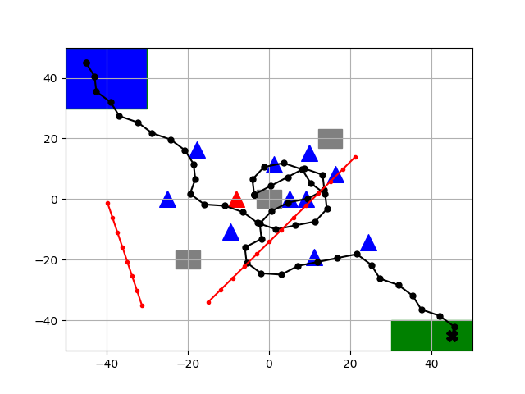}
			\subcaption{\scriptsize HST-strategy}
		\end{minipage}
		\caption{\small Examples of typical UAV trajectory in different scenarios. }
		\label{Fig:fix_jammer_traj}
	\end{figure*}

	Fig. \ref{Fig:fix_jammer_traj} presents the UAV trajectories in no-jammer, continuous jamming attack when $P^J_{l} =10^{-3}/3$W (with no defense), virtual jammer defense strategy (VJ-strategy), and higher SINR threshold defense strategy (HST-strategy) scenarios.  Fig. \ref{Fig:fix_jammer_traj}(a) shows that the typical UAV can find an efficient trajectory to complete its mission when there is no jammer. Fig. \ref{Fig:fix_jammer_traj}(b) shows that the typical UAV trajectory becomes curvy (with several loops) due to the existence of the jammer at (-8,0). Figs. \ref{Fig:fix_jammer_traj}(c) and (d) demonstrate that with the two defense strategies, the typical UAV is able to complete its mission in shorter trajectories under continuous  jamming attacks, while the trajectories are still relatively longer than that in the no-jammer scenario.
	
	\subsection{Periodic Jamming Attack Scenario}
\begin{figure*}
		\centering
		\begin{minipage}{0.32\textwidth}
			\centering
			\includegraphics[width=1\textwidth]{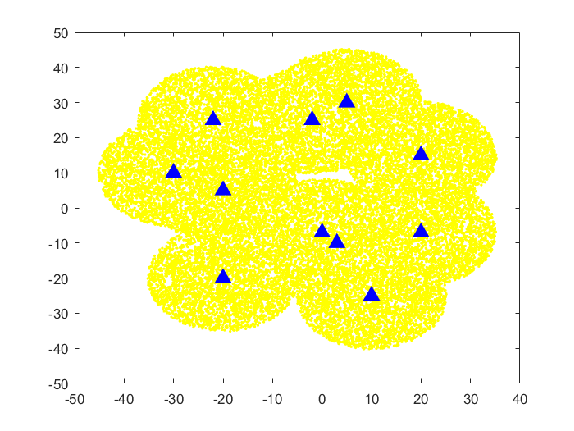}
			\subcaption{\scriptsize No jammer }
		\end{minipage}
		\begin{minipage}{0.32\textwidth}
			\centering
			\includegraphics[width=1\textwidth]{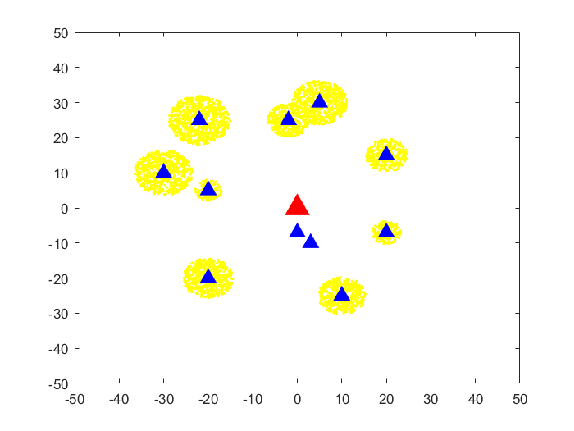}
			\subcaption{\scriptsize $P^J_{h1} = 10^{-3}/2.5$W}
		\end{minipage}
		\begin{minipage}{0.32\textwidth}
			\centering
			\includegraphics[width=1\textwidth]{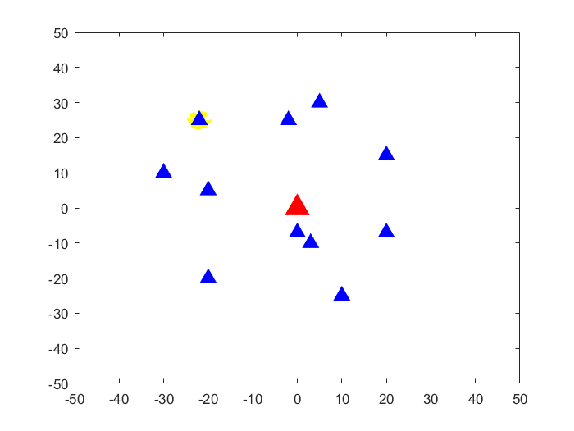}
			\subcaption{\scriptsize $P^J_{h2} = 10^{-3}/2$W}
		\end{minipage}
		\caption{\small Illustrations of the reliable transmission region when the jammer is absent or is located at (0,0) with different transmit powers. }
		\label{Fig:fix_jammer_higher_power}
	\end{figure*}

	In this subsection, it is assumed that the jammer interferes periodically with transmit power $P^J_{h}$ and jamming period $\tau^J_{h}$. We adopt two  transmit power levels $P^J_{h1} = 10^{-3}/2.5$W and $P^J_{h2} = 10^{-3}/2$W. Since  $P^J_{l} \times 60 = P^J_{h} \times \tau^J_{h}$, we have $ \tau^J_{h1}= \frac{P^J_{l}\times 60}{P^J_{h1}} = 50$s and $ \tau^J_{h2} = 40$s. In other words, the jammer transmits with $P^J_{h1}$ for 50s  and becomes silent for 10s per minute, or transmits with $P^J_{h2}$ for 40s and becomes silent for 20s per minute.
	Fig. \ref{Fig:fix_jammer_higher_power} illustrates examples of the reliable transmission region when the jammer has different transmit powers. As we see from the figures, the larger the transmit power is, the greater influence the jammer exerts, e.g., when  $P^2_{Jh} = 10^{-3}/2$W, majority of the connections are blocked during jamming.  Table \ref{Table:period_jammer_attack} provides the performances of the periodic jamming attack and two aforementioned defense strategies. The results in the table indicate that the influence of the periodic jamming attack is not significant, due to the reason that the typical UAV is able to wait for the jammer to become silent and then collect data from the IoT nodes. However, overall the SR and TR are still reduced due to the longer mission completion time caused by waiting. Overall, the collision rate is under 0.6\%, thus is not listed in the table. Using the proposed defense strategies, the performance can again be recovered to levels close to those in the  no-jammer scenario.
	
	\begin{table}[h]
		\centering
		\caption{Periodic jamming attack performance and defense performance.}
		\label{Table:period_jammer_attack}
		\begin{tabular}{c|c|c|c|c|c}
			\hline\hline
			\multicolumn{2}{c|}{}                        & SR(\%)   & DR(\%)   & TR(\%)    & Reward \\ \hline
			\multicolumn{2}{c|}{No-Jammer}               & 99.4 & 100  & 100 &  81.904 \\ \hline
			\multirow{2}{*}{Attack}  & $P^J= \frac{10^{-3}}{2}$    & 99.1 & 99.3 & 99.5 & 72.355 \\ \cline{2-6}
			& $P^J= \frac{10^{-3}}{2.5}$ & 96.9 & 99.3 & 97.3  & 50.257 \\ \hline
			\multirow{2}{*}{Defense} & $P^J= \frac{10^{-3}}{2}$   & 99.6 & 100  & 99.9 &   74.269 \\ \cline{2-6}
			& $P^J= \frac{10^{-3}}{2.5}$ & 99.1 & 100  & 99.8  & 63.151 \\ \hline \hline
		\end{tabular}
	\end{table}

	\subsection{Intelligent Jamming Attack Scenarios}
	
	\subsubsection{Original Policies of the Typical UAV}
	The typical UAV has two original policies learned in the no-jammer scenarios for different mission completion deadlines $\mT^V_t$, and these policies are denoted as
	\begin{itemize}
		\item $\pi^V_1$ when $\mT^V_t=100$s;
		\item $\pi^V_2$ when $\mT^V_t=200$s.
	\end{itemize}
	The performances of these two original policies in a no-jammer scenario are provided in Table \ref{Table:intelligent_jammer_nojammer}. From the table, we observe that the SR is at least 98.9\% and DR is close to 100\%, indicating that the typical UAV can complete its mission with 98\% success rate if no jammer exists. In addition, with the looser mission completion deadline of $\mT^V_t = 200$s, the overall performance can be increased. Note that since the reward function is modified within the defense algorithm, we do not compare the reward performances in this section.
	\begin{table}[h]
		\centering
		\caption{Performance of typical UAV's policies in the absence of jamming attacks.}
		\label{Table:intelligent_jammer_nojammer}
	\begin{tabular}{c|c|c|c|c}
		\hline \hline
		& SR(\%) & DR(\%) & TR(\%) & CR(\%) \\ \hline
		$\pi^V_1$	& 98.9   & 99.9   & 99.4    & 0.4          \\ \hline
		$\pi^V_2$	& 99.4   & 100   & 100      & 0.6      \\ \hline\hline
	\end{tabular}
	\end{table}
	
	\subsubsection{Intelligent Jamming Attack Performance}
	In this subsection, the jammer flies at height $H_J=30$m with transmit power $P^J = 10^{-3}/3$W or $10^{-4}$W.
	Four jammers are trained to attack the typical UAV with different transmit power levels $P^J$, and these jammers are described in more detail below:
	\begin{itemize}
		\item Jammer 1 (J1) is trained to attack $\pi^V_1$ with transmit power $P^J = 10^{-3}/3$W, and its policy is denoted by $\pi^J_1$;
		\item  Jammer 2 (J2) is trained to attack $\pi^V_2$ with transmit power $P^J = 10^{-3}/3$W, and its policy is denoted by $\pi^J_2$;
		\item Jammer 3 (J3) is trained to attack $\pi^V_1$ with transmit power $P^J = 10^{-4}$W, and its policy is denoted by $\pi^J_3$;
		\item Jammer 4 (J4) is trained to attack $\pi^V_2$ with transmit power $P^J = 10^{-4}$W, and its policy is denoted by $\pi^J_4$.
	\end{itemize}
	The attack performances of the jammers  are provided in Table \ref{Table:intelligent_jammer_attack}. By comparing the typical UAV's  SR, DR, and TR in the no-jammer scenario (provided in Table \ref{Table:intelligent_jammer_nojammer}) and in the presence of different jammers (provided in Table \ref{Table:intelligent_jammer_attack}), we observe that each jammer can significantly reduce the SR, DR and TR. The substantial decrease in DR is due to the reason that the jammer is encouraged to get close to the typical UAV and thus interference from the jammer can be very large, leading to the result that most connections are blocked and the typical UAV fails to collect data from some nodes. This is also the reason that the DR in the presence of jammers J1 and J2 (where $P^J = 10^{-3}/3$W)  is much smaller than DR in the presence of jammers J3 and J4 (where $P^J = 10^{-4}$W). In addition, the SR and TR decrease due to the following two reasons: 1) the reliable transmission region is substantially reduced with the existence of the jammer, and thus the typical UAV needs more time to collect data from the nodes; and 2) the reliable transmission region is dynamically changing due to the movement of the jammer, and that leads the typical UAV not to choose the optimal action and generally need more time to arrive at its destination, and therefore violating its mission completion deadline.  These are also the reasons for why SR and TR when $\mT^V_t = 100$s (with jammers J1 and J3 in Table \ref{Table:intelligent_jammer_attack}) are smaller than  SR and TR when $\mT^V_t = 200$s (with jammers J2 and J4 in Table \ref{Table:intelligent_jammer_attack}).	
	\begin{table}[h]
		\centering
		\caption{Performance of typical UAV in the presence  of different jammers.}
		\label{Table:intelligent_jammer_attack}
	\begin{tabular}{c|c|c|c|c}
		\hline \hline
		Jammer& SR(\%) & DR(\%) & TR(\%) & CR(\%)  \\ \hline
		J1	& 0.7    & 13.8    & 1.3   & 0.7        \\ \hline
		J2	& 7.6  & 5.3   & 8.1     & 0.5           \\ \hline
		J3	& 2.1   & 49.8   & 2.6    & 0.5          \\ \hline
		J4	& 33.7   & 77.8    & 34.4     & 0.7          \\ \hline\hline
	\end{tabular}
	\end{table}

	\subsubsection{Defense Performance}
	Using the proposed defense algorithm with updated state space  and reward function, policies can be re-trained against the intelligent jammers. To defend against the jammers designed in the previous subsection, we re-train the typical UAV's policy. The re-trained  policies are listed and described below:
	\begin{itemize}
		\item $\pi^{Vd}_1$ is trained with the existence of J1, i.e., $\pi^{Ud}_1$ is trained to defend against J1;
		\item $\pi^{Vd}_2$ is trained to defend against J2;
		\item $\pi^{Vd}_3$ is trained to defend against J3;
		\item $\pi^{Vd}_4$ is trained to defend against J4.
	\end{itemize}
	Since the typical UAV needs more time to finish its mission due to the significant reduction in the reliable transmission region, we loosen the mission completion deadline $\mT^V_t$ in defensive strategies. The performances of defense polices are provided in Table \ref{Table:intelligent_jammer_defend}.	From the rows in boldface (in which we have the performance results of the retrained policies against the corresponding jamming attackers), we observe that the performance of the typical UAV in terms of SR, DR and TR is considerably restored. More specifically, the DR is recovered to above 80\% when defending against J1, J3, and J4, and above 70\% when defending against J2. Also, the SR and TR are recovered to above 94\%. The reasons for this significant improvement are the following: 1) with loosened mission completion deadline, the typical UAV is allowed to use more time  to  collect data from the IoT nodes; 2) with the presence of the jammer in training, the UAV learns the dynamically varying reliable transmission regions; and 3) the typical UAV's policy is updated and re-trained, and thus the jammer cannot predict the typical UAV's movement well.
	
	In addition, we also use the policies $\pi^{Vd}_2$ and $\pi^{Vd}_4$ to defend against other jammers (with respect to which the defensive policies have not been retrained). It is observed that both policies can recover the performance to some extent, especially when using $\pi^{Vd}_2$. Even though the performances are generally not as good as the case of using the matching defense policy, the SR and TR are above 80\% and DR is above 70\%, which are much higher than the performance without any defense.
	\begin{table}[h]
		\centering
		\caption{Performances of defense policies in the intelligent jamming attack scenarios.}
		\label{Table:intelligent_jammer_defend}
	\begin{tabular}{c|c|c|c|c|c}
		\hline \hline
		&Jammer & SR(\%) & DR(\%) & TR(\%) & CR(\%)  \\ \hline
		$\pi^{Vd}_1$ &\textbf{J1}	&\textbf{94.7} &\textbf{87.7} &\textbf{95.6} &\textbf{0.9}   \\ \hline
		\multirow{4}{*}{$\pi^{Vd}_2$}
		& J1 & 94.8    & 82.1    &94.6    & 1.1          \\ \cline{2-6}
		& \textbf{J2} &\textbf{94.8} &\textbf{71.5} &\textbf{95.4} &\textbf{0.6}        \\ \cline{2-6}
		& J3 & 98.8   &89.2    & 99.4    &0.6           \\ \cline{2-6}
		& J4 & 96.9    & 78.4    & 98.1    & 1.1           \\ \hline
		$\pi^{Vd}_3$ &\textbf{J3}	&\textbf{98.1} &\textbf{84.6} &\textbf{98.5} &\textbf{0.3}        \\ \hline
		\multirow{4}{*}{$\pi^{Vd}_4$}
		& J1  & 83   &   88.3  & 83.9    &0.9          \\ \cline{2-6}
		& J2  & 81.9   & 79.8   &82.7    &0.8           \\ \cline{2-6}
		& J3 &87.2    & 84.1   & 88.2   & 0.9         \\ \cline{2-6}
		& \textbf{J4} &\textbf{98.4} &\textbf{82.3} &\textbf{99.1} &\textbf{0.7}       \\ \hline \hline
	\end{tabular}
	\end{table}

\begin{figure}
		\centering
		\includegraphics[width=0.45\textwidth]{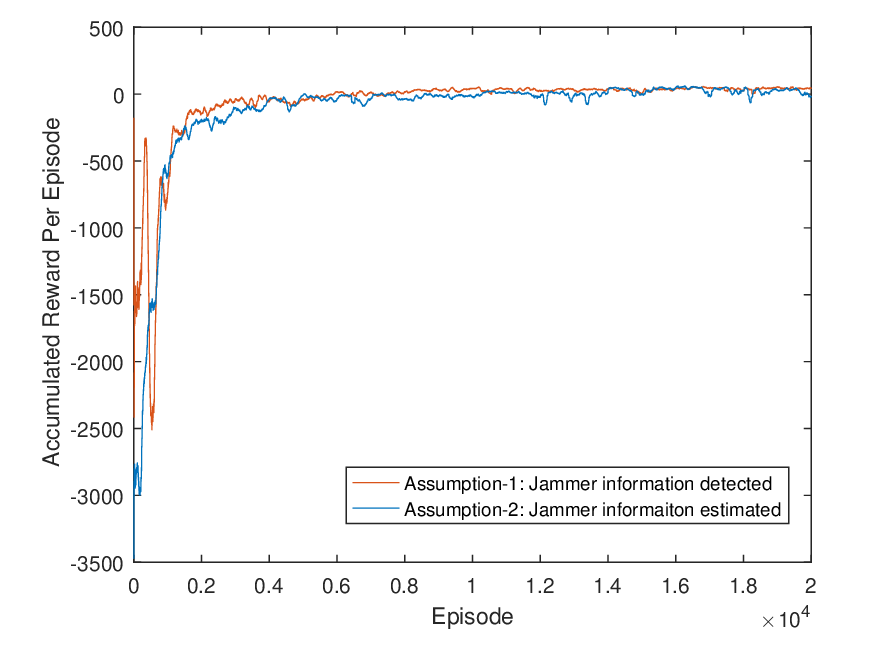}
		\caption{\small Comparison of accumulated reward per episode between two scenarios. In the scenario with assumption-1, the jammer's information can be detected all the time, while in the scenario with assumption-2, the jammer's information is estimated when it is outside of $\mathbb{O}$.}
		\label{Fig:assumption_comparison}
		\end{figure}

	The proposed algorithm can be extended and utilized in more realistic scenarios, e.g., in a scenario in which the typical UAV is not able to detect the jammer all the time. Particularly, if the jammer is in sensing region $\mathbb{O}$, its position, velocity and orientation can be sensed. Otherwise, the typical UAV fails to sense this information. In this more practical setting, in order to predict the jammer's information, a velocity filter is designed to obtain the estimated next velocity $\hat{\vvec}^J_{t+1}$ using the jammer's velocities in the past $\tau$ time steps, and the estimated velocity can now be expressed as
	\begin{align}
	\hat{\vvec}^J_{t+1} = \frac{1}{\tau} \sum_{t'=t-\tau}^{t} \vvec^J_{t'},  \qquad \text{if $\pvec^J_{t+1} \notin \mathbb{O}$}
	\end{align}
	where $\pvec^J_{t+1}$ is the jammer's position  at time step $t+1$. Then, the next estimated position is $\hat{\pvec}^J_{t+1} = \pvec^J_{t} + \hat{\vvec}^J_{t+1}\times \Delta t $ and the next estimated orientation is $\hat{\phi}_{t+1} = \arctan \hat{v}^J_{t+1,y}/\hat{v}^J_{t+1,x}$. Fig. \ref{Fig:assumption_comparison} plots the reward values in training when considering the scenario with assumption-1 (in which the jammer's information is detected all the time) and the scenario with assumption-2 (in which the jammer's information is estimated using the velocity filter if it is outside $\mathbb{O}$). From Fig. \ref{Fig:assumption_comparison}, we observe that we can achieve comparable performance in the scenario with assumption-2 compared to that with assumption-1. This follows from two reasons. First, due to the kinematic constraints, sudden drastic changes in the jammer's velocity are not allowed, and thus the past movements provide relatively accurate indications on its near-term future mobility. Secondly,  if the jammer is outside of the typical UAV's sensing region $\mathbb{O}$, it is  far away from the typical UAV, and correspondingly its interference is small, leading to small influence on the typical UAV's transmission. Thus, the estimation error does not impact the typical UAV's performance substantially.

\begin{figure}[h]
		\centering
		\includegraphics[width=0.45\textwidth]{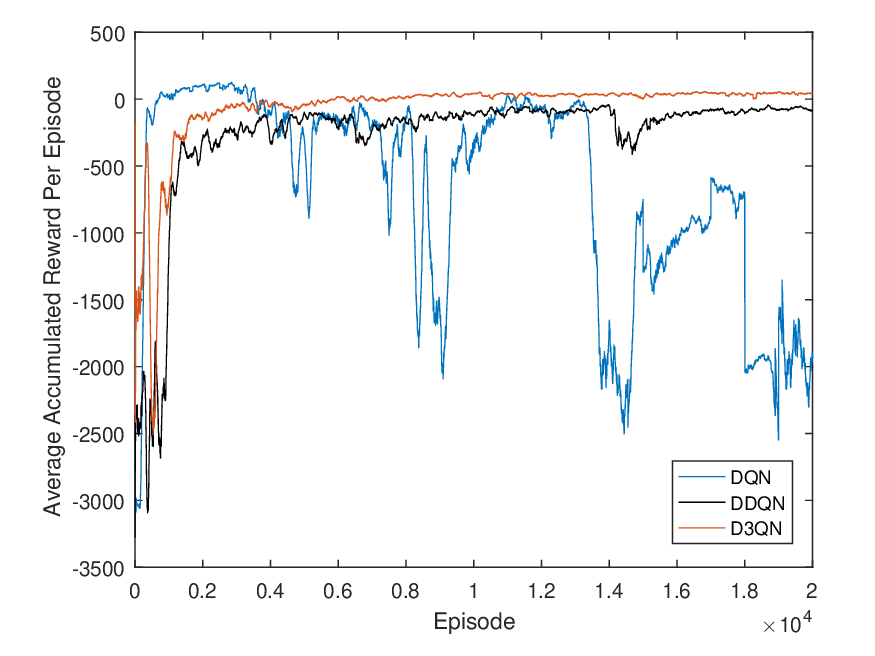}
		\caption{\small Comparison of accumulated reward per episode among DQN, DDQN and D3QN. }
		\label{Fig:converges_comparison}
	\end{figure}

	In the literature, Q-learning (e.g., \cite{jam_li2018protecting},\cite{jam_peng2019anti} ) and DQN (e.g., \cite{jam_xiao2017user}, \cite{jam_gao2019anti}) have been used to defend against jamming attacks. Due to the large size of the state space, Q-learning is typically infeasible to be used in such studies.  Figure \ref{Fig:converges_comparison} depicts the reward when utilizing DQN, DDQN and D3QN to train the typical UAV's defense policy. It can be observed that D3QN is more rewarding  and converges much faster. Since the dueling architecture is able to learn which states are valuable without learning the effect of each action for each state, it has the ability to identify the correct action more quickly  during policy evaluation \cite{wang2016dueling}.

	\begin{figure*}
	\centering
	\begin{minipage}{0.32\textwidth}
		\centering
		\includegraphics[width=1\textwidth]{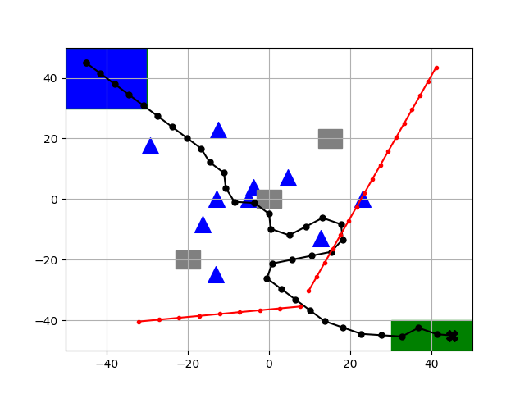}
		\subcaption{\scriptsize No jammer }
	\end{minipage}
	\begin{minipage}{0.32\textwidth}
		\centering
		\includegraphics[width=1\textwidth]{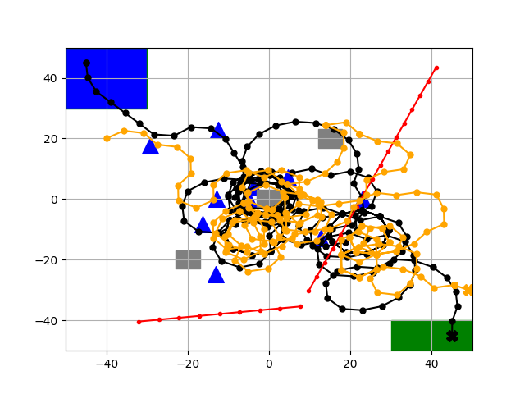}
		\subcaption{\scriptsize Intelligent jamming attack (with J2)}
	\end{minipage}
	\begin{minipage}{0.32\textwidth}
		\centering
		\includegraphics[width=1\textwidth]{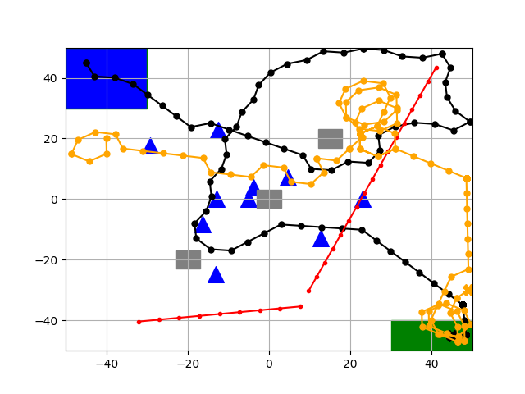}
		\subcaption{\scriptsize Defense (with $\pi^{Vd}_2$)}
	\end{minipage}
	\caption{\small Examples of typical UAV trajectory in different scenarios. }
	\label{Fig:intelligent_jammer_traj}
\end{figure*}

	\subsubsection{Trajectory Designs}
	We provide examples of UAV trajectories in Fig. \ref{Fig:intelligent_jammer_traj} in the no-jammer scenario, in the scenario with intelligent jamming attack and no defense, and in the scenario in which defensive policy is employed.  Note that the orange-dotted lines are the intelligent jammer's trajectories. In Fig. \ref{Fig:intelligent_jammer_traj}(b), we observe that the intelligent jammer can follow the typical UAV closely if the UAV does not implement the defense policy, and the jammer makes the typical UAV trajectory really curvy and long (compared with the trajectory in  Fig. \ref{Fig:intelligent_jammer_traj}(a) where no jammer exists). In addition, Fig. \ref{Fig:intelligent_jammer_traj}(c) shows that if the defense strategy is utilized and the typical UAV's policy is updated, the intelligent jammer is not able to follow the UAV well. Therefore, the typical UAV can find  an efficient trajectory to complete its mission (e.g., a short trajectory not exceeding the mission completion deadline, and being close to the IoT nodes but away from the jammer in order to collect data). This observation further verifies the effectiveness of the proposed and implemented defensive measures.

	\section{Conclusion}
	In this paper, we have investigated jamming-resilient UAV path planning strategies for data collection in IoT networks, in which the typical UAV can learn the optimal trajectory to elude such jamming attacks.  Specifically, the typical UAV is required to collect data from multiple distributed IoT nodes under collision avoidance, mission completion deadline, and kinematic constraints in the presence of jamming attacks.  We have first designed a fixed ground jammer with continuous jamming attack and periodic jamming attack strategies to inject interference into the link between the typical UAV and IoT nodes. RL-based defensive strategies that utilize a virtual jammer and adopt a higher SINR threshold are proposed against these attacks.  Secondly, we have designed an intelligent UAV jammer, which uses an RL algorithm to choose actions based on its observation.  Finally,  an intelligent UAV anti-jamming strategy is developed to defend against such intelligent jamming attacks.   The optimal trajectory of the typical UAV is obtained via D3QN. Simulation results have shown that both fixed jamming and intelligent UAV jamming attacks have significant influence on the typical UAV's performance, and the proposed defense strategies can recover the performance close to that in the no-jammer scenario.

	\bibliographystyle{IEEEtran}
	\bibliography{compresensive2}


\end{document}